\documentclass[pdflatex,sn-basic,iicol]{sn-jnl}
\usepackage{anyfontsize}
\usepackage{amsmath,amssymb,amsfonts}
\usepackage{amsthm}
\usepackage{graphicx} 
\usepackage{acronym}
\usepackage{manyfoot}
\usepackage[dvipsnames]{xcolor}
\usepackage{physics}
\usepackage{manyfoot}%
\usepackage{hyperref}
\usepackage[capitalise]{cleveref}
\usepackage{caption}

\usepackage{subcaption}
\usepackage[normalem]{ulem}

\acrodef{ai}[AI]{Artificial Intelligence}
\acrodef{ann}[ANN]{Artificial Neural Network}
\acrodef{gen-ai}[GenAI]{Generative Artificial Intelligence}
\acrodef{ml}[ML]{Machine Learning}
\acrodef{dl}[DL]{Deep Learning}
\acrodef{fid}[FID]{Fr\'echet inception distance}
\acrodef{mlp}[MLP]{Multilayer Perceptron}
\acrodef{relu}[ReLU]{Rectified Linear Unit}
\acrodef{pmf}[PMF]{Probability Mass Function}

\acrodef{kl}[KL]{Kullback–Leibler}

\acrodef{crw}[CRW]{Classical Random Walk}

\acrodef{nf}[NF]{Normalizing Flow}
\acrodef{vae}[VAE]{Variational Auto-Encoder}
\acrodef{gan}[GAN]{Generative Adversarial Network}

\acrodef{dm}[DM]{Diffusion Model}
\acrodef{ddpm}[DDPM]{denoising diffusion probabilistic model}

\acrodef{nisq}[NISQ]{Noisy Intermediate-Scale Quantum}
\acrodef{qpu}[QPU]{Quantum Processing Unit}
\acrodef{qw}[QW]{Quantum Walk}
\acrodef{dtqw}[DTQW]{Discrete-time Quantum Walk}
\acrodef{ctqw}[CTQW]{Continuous-time Quantum Walk}
\acrodef{qsw}[QSW]{Quantum Stochastic Walk}

\acrodef{q-ai}[QAI]{Quantum Artificial Intelligence}
\acrodef{q-gen-ai}[QGenAI]{Quantum Generative Artificial Intelligence}
\acrodef{qml}[QML]{Quantum Machine Learning}

\acrodef{qgan}[QGAN]{Quantum Generative Adversarial Network}
\acrodef{qdm}[QDM]{Quantum Diffusion Model}
\acrodef{dsqdm}[DSQDM]{Discrete State-space Quantum Diffusion Model}

\newcommand{\ve}[1]{\ensuremath{\mathbf{#1}}} 
\newcommand{\change}[1]{#1}

\begin{document}

\title{Physics-inspired Generative AI models via real hardware-based noisy quantum diffusion}

\author[1]{\fontsize{11.91pt}{14.3pt}\selectfont\fnm{Marco} \sur{Parigi}}\email{marco.parigi@unifi.it}

\author*[1,2]{\fontsize{11.91pt}{14.3pt}\selectfont\fnm{Stefano} \sur{Martina}}\email{stefano.martina@unifi.it}

\author[1,3,4]{\fontsize{11.91pt}{14.3pt}\selectfont\fnm{Francesco Aldo} \sur{Venturelli}}\email{francescoaldo.venturelli@upf.edu}

\author[1,2]{\fontsize{11.91pt}{14.3pt}\selectfont\fnm{Filippo} \sur{Caruso}}\email{filippo.caruso@unifi.it}

\affil[1]{\orgdiv{Department of Physics and Astronomy}, \orgname{University of Florence}, \orgaddress{\street{Via Sansone 1}, \city{Sesto Fiorentino}, \postcode{50019}, \state{Florence}, \country{Italy}}}

\affil[2]{\orgdiv{LENS - European Laboratory for Non-Linear Spectroscopy}, \orgname{University of Florence}, \orgaddress{\street{Via Nello Carrara 1}, \city{Sesto Fiorentino}, \postcode{50019}, \state{Florence}, \country{Italy}}}

\affil[3]{\orgdiv{Department of Engineering}, \orgname{University Pompeu Fabra}, \orgaddress{\street{Tànger 122-140}, \city{Barcelona}, \postcode{08018}, \state{Barcelona}, \country{Spain}}}

\affil[4]{\orgdiv{Quantic}, \orgname{Barcelona Supercomputing Center}, \orgaddress{\street{Plaça Eusebi Güell 1-3}, \city{Barcelona}, \postcode{08034}, \state{Barcelona}, \country{Spain}}}

\abstract{Quantum Diffusion Models (QDMs) are an emerging paradigm in Generative AI that aims to use quantum properties to improve the performances of their classical counterparts. However, existing algorithms are not easily scalable due to the limitations of near-term quantum devices. Following our previous work on QDMs, here we propose and implement two physics-inspired protocols. In the first, we use the formalism of quantum stochastic walks, showing that a specific interplay of quantum and classical dynamics in the forward process produces statistically more robust models generating sets of MNIST images with lower Fr\'echet Inception Distance (FID) than using totally classical dynamics. In the second approach, we realize an algorithm to generate images by exploiting the intrinsic noise of real IBM quantum hardware with only four qubits. Our work could be a starting point to pave the way for new scenarios for large-scale algorithms in quantum Generative AI, where quantum noise is neither mitigated nor corrected, but instead exploited as a useful resource.}


\maketitle

\section{Introduction}
\ac{gen-ai} is one of the most interesting recent research fields that uses \ac{ml} models capable of learning the underlying structure from a finite set of samples to create new, original, and meaningful content, such as images, text, or other forms of data. Nowadays, \ac{gen-ai} technology is used in both academic and industrial applications to find new, creative, and efficient solutions to real-life problems. Over the years, different generative models have been proposed, such as \acp{gan}~\citep{Goodfellow2020}, \acp{vae}~\citep{Kingma2019}, and \acp{nf}~\citep{Rezende15NormFlows}, which have shown great success in generating high-quality novel data. However, Denoising Probabilistic Diffusion Models~\citep{Sohl2015, Ho2020} (or simply \acp{dm}) have recently achieved state-of-the-art performance by overcoming previous models in generative tasks, such as image and audio synthesis~\citep{Paiano_2024, Dhariwal2021, kongdiffwave2021}. \acp{dm} have been introduced by~\cite{Sohl2015} and are inspired by the physical phenomenon of non-equilibrium thermodynamics, i.e., \emph{diffusion}. The generic pipeline of \acp{dm} consists of two Markov chains that are called forward (or diffusion) and backward (or denoising). In the forward chain, classical noise is injected through a stochastic process into the training samples until they become totally noisy. In the backward chain, \acp{ann} are trained to iteratively remove the aforementioned perturbation to reverse the forward process in order to learn the unknown distribution of the training samples and thus generate new samples. Currently, \acp{dm} are widely adopted in computer vision tasks~\citep{saharia2022photorealistic, Rombach_2022_CVPR, Lugmayr_2022_CVPR}, text generation~\citep{austin2021structured}, sequential data modeling~\citep{Tashiro2021}, audio synthesis~\citep{kongdiffwave2021}, and are one of the fundamental elements of famous and widespread \ac{gen-ai} technologies, such as Stable Diffusion~\citep{stableDiffusion}, DALL-E 4~\citep{dalle4}, and Diffwave~\citep{kongdiffwave2021}.

On the other hand, quantum computing is a rapidly emerging technology that harnesses peculiar quantum mechanical phenomena such as \emph{superposition}, \emph{entanglement}, and \emph{coherence} to solve complex problems with fewer resources, or tasks that are intractable with classical (super)computers. For instance, milestone quantum algorithms such as Shor's factorization~\citep{Shor1994algorithms, ShorPolyTime} and Grover's search~\citep{Grover1996} have exponential and quadratic \emph{speed-ups}, respectively, over their classical counterparts. Moreover, quantum computing promises to achieve a speed-up in simulating quantum systems~\citep{Georgescu2014, OMalley2016, Babbush2018}, solving linear systems of equations~\citep{Harrow2009}, and optimization tasks~\citep{Crosson2016, farhi2019}. However, these algorithms require \emph{fault-tolerant} quantum processors~\citep{preskill1998fault}, i.e.\ hardware with a large number of error-corrected qubits. Consequently, they are not feasible on currently available \ac{nisq} devices~\citep{preskill2018nisq} that use \acp{qpu}~\citep{IBMqpu, queraQPU} composed of a few hundred qubits highly prone to quantum noise. 
In order to reduce the effects of noise, quantum error correction techniques can be applied, which, however, require an elevated number of physical qubits~\citep{Knill1997, Campbell2024}. In this context, Google Quantum AI has recently developed a new quantum device called \emph{Willow} that seems to allow for an exponential reduction of errors while increasing the number of qubits~\citep{Acharya2024Willow}.

An active area of research in quantum computation involves algorithms based on \acp{qw}. \acp{qw} have been formally introduced by~\cite{Aharonov1993qw} as the quantum mechanical counterpart of \acp{crw} and are exploited in many quantum protocols today. It has been shown that \acp{qw} outperform \acp{crw}, for example, in search algorithms~\citep{Childs2004, Ambainis2005, Magniez2011}, transport phenomena~\citep{Caruso2014, Caruso2016, DallaPozza2022}, secure communications and cryptography protocols~\citep{AbdElLatif2020, Zeuner2021}, and distinguishing hard graphs~\citep{kasture2025multiqw}. \acp{qw} can also be used as a primitive of a universal model for quantum computation~\citep{Lovett2010, Singh2021, Chawla2023} and can be\emph{efficiently} implemented by physical experiments~\citep{Dur2002, Schreiber2012, Goyal2013} and quantum processors~\citep{Lahini2018, Acasiete2020, Razzoli2024}.

\acp{qw} are part of a more general family: \acp{qsw}~\citep{whitfield2010quantum}, allowing one to describe the evolution of a quantum mechanical walker by means of quantum stochastic equations of motion and also to generalize classical random walkers.

\ac{qml} is an emerging field that integrates quantum computing and \ac{ml} techniques~\citep{wittek2014quantum, biamonte2017quantum,schuld2021qml}. However, due to the limitations of \ac{nisq} devices, many \ac{qml} algorithms are usually applied to toy problems where data are reduced in terms of the number of features or integrated with classical models implementing the so-called \emph{hybrid} quantum-classical algorithms (or \ac{nisq} algorithms)~\citep{McClean2016, Bharti2022}. Recently, a plethora of these algorithms have been proposed in different applications, for example: image processing~\citep{Das2023, Geng2022, parigi2025supervised}, quantum chemistry~\citep{Peruzzo2014, Kandala2017}, combinatorial and optimization problems~\citep{Gonzalez2022}, searching algorithms~\citep{Zhang2021}, and machine learning tasks~\citep{farhi2018}.
In the context of \ac{q-gen-ai}, previous works generalize classical \ac{gen-ai} models into the quantum domain: \acp{qgan}~\citep{Lloyd2018, Dallaire-Demers2018}, QVAE~\citep{Khoshaman2019}, and \acp{qdm}~\citep{Parigi2024}. An interesting aspect of \ac{q-gen-ai} models is that they allow the integration of computational protocols with physical quantum devices. For example, \acp{qgan} have recently been realized using a silicon quantum photonic chip~\citep{ma2024qgan_on_chip}. Moreover, a practical \emph{quantum advantage} in generative modeling has been demonstrated in the data-limited scenario, comparing quantum-against-classical generative algorithms~\citep{Hibat-Allah2024}. 

Concerning \acp{qdm}, numerical simulations show how the design of quantum-classical algorithms can improve the quality of the generated data~\citep{kölle2024quantumdenoisingdiffusionmodels, DeFalco2024}, learn complex quantum distributions~\citep{cacioppo2024quantum2}, reduce the number of trainable parameters in the denoising \ac{ann}~\citep{DeFalco2024RedPar}, and potentially achieve sampling speeds faster than those of classical \acp{dm}~\citep{kivijervi2024quantum}. However, one of the main challenges of these algorithms is their \emph{scalability} on near-term quantum processors. In fact, the currently proposed \acp{qdm} approaches are usually implemented in simplified scenarios~\citep{Parigi2024, zhang2024generative, chen2024quantumgenerativediffusionmodel}, or by using pre-processing techniques~\citep{kölle2024quantumdenoisingdiffusionmodels} and classical latent models~\citep{cacioppo2023quantumdiffusionmodels} to reduce the dimensional representation of the data. In addition, the idea of harnessing quantum noise to corrupt data in the forward diffusion process has been explored in simulated scenarios through quantum noise channels~\citep{Parigi2024, chen2024quantumgenerativediffusionmodel, kwun2024mixedstatequantumdenoisingdiffusion}.

In this work, we first study the performances in image generation of \change{unconditional} \acp{dm} when classical diffusion is replaced or integrated with quantum stochastic dynamics in the forward process. \change{The backward process is realized classically}. In particular, using the formalism of \acp{qsw}, we show that a specific interplay of quantum-classical stochastic dynamics improves image generation quality, leading to lower \ac{fid} values between real and generated samples. The hybrid model is also statistically more robust than the classical \acp{dm}. Then, in the second part, we implement a \ac{qw} dynamics by using a quantum circuit and exploit the intrinsic noise of a real IBM quantum processor to generate images from the MNIST dataset~\citep{lecun2010mnist}.

\section{Methods}\label{sec:methods}
\subsection{Diffusion Models}
\acfp{dm} are a class of latent variable generative models trained to learn the underlying unknown data distribution of a finite dataset in order to generate new similarly distributed synthetic data. The core idea is to use a classical Markov chain to gradually convert an \change{initial unknown data distribution $q(\ve{x}_0)$}, called the \emph{posterior}, to a simple, well-known distribution, e.g., Gaussian or uniform, called the \emph{prior}.
The most generic pipeline of \acp{dm} is characterized by a forward (or diffusion) and a backward (or denoising) process.
In the forward process, an initial sample $\ve{x}_0 \sim q(\ve{x}_0)$ is corrupted in a sequence of $T$ increasingly noisy latent variables $\ve{x}_1, \ve{x}_2, \dots, \ve{x}_T$. Formally, the diffusion process is described by a classical random process via a Markov chain that gradually injects noise into the initial sample $\ve{x}_0$. In this process, the features of the samples are mathematically represented as classical random walkers undergoing stochastic dynamics~\citep{Ho2020}. More precisely, this is realized by using the Markov transition kernel:
\begin{equation}\label{eq: transition_kernel}
    q(\ve{x}_{t} | \ve{x}_{t-1}) = K_{\pi}(\ve{x}_{t}| \ve{x}_{t-1}; \beta_t),
\end{equation}
\begin{equation}
    q(\ve{x}) = \int d\ve{x}'\ K_{\pi}(\ve{x}| \ve{x}'; \beta) q(\ve{x}'),
\end{equation}
where $\beta_{t} \in (0, 1)$ is a hyper-parameter at time $t$ of the model (or in physical terms, the diffusion rate) describing the \emph{level of noise} added at each time step, and $\ve{x}_{t}$ and $\ve{x}_{t-1}$ are the random noisy latent variables at the time steps $t$ and $t-1$, respectively.
The scheduling of $\beta_t$ is chosen and fixed such that the initial data distribution $q(\ve{x}_0)$ converges to a well-known stationary distribution $\pi(\ve{x}_T)$ in the limit $T \rightarrow \infty$.
The forward trajectory after performing $T$ time steps of diffusion can be written as:
\begin{equation}
    q(\ve{x}_{0:T})= \change{q(\ve{x}_0)}\prod_{t=1}^T q(\ve{x}_{t} | \ve{x}_{t-1}),
\end{equation}
where the chain rule of probability and the Markov property of the process are used to factorize the joint distribution $q(\ve{x}_{0:T})$. In addition, the diffusion process contains no trainable parameters and does not involve any learning model.
The idea of the backward process is to reverse the forward dynamics, moving from a pure noise sample $\ve{x}_T \sim \pi(\ve{x}_T)$ towards a sample of the initial distribution $q(\ve{x}_0)$. The denoising is implemented by an \ac{ann} that is trained to learn the reverse trajectory of the diffusion process: 
\begin{equation}
    p(\ve{x}_T) = \pi(\ve{x}_T)
\end{equation}
\begin{equation}
    p_\theta(\ve{x}_{0:T})=p(\ve{x}_T)\prod_{t=1}^T p_\theta(\ve{x}_{t-1} | \ve{x}_{t}),
\end{equation}
where $p_\theta(\ve{x}_{t-1} | \ve{x}_{t})$ is a parameterized transition kernel having the same functional form of the forward kernel $q(\ve{x}_{t} | \ve{x}_{t-1})$. 
A deep \ac{ann}, usually with a U-Net architecture~\citep{Ronneberger2015}, is used to estimate the parameters $\theta$ at each time step.
The denoising network is trained to optimize the negative log-likelihood on the training data writing the evidence lower bound (ELBO)~\citep{Kingma2014} as follows:
\begin{equation}\label{eq: ELBO_bound}
\begin{split}
    L & = D_\text{KL}(q(\ve{x}_{0:T})|| p_\theta(\ve{x}_{0:T}))\\
      & = - \mathbb{E}_q [\log p_\theta(\ve{x}_{0:T})] + \text{const} \\
      & = \mathbb{E}_q \Bigg[-\log p(\ve{x}_T) - \change{\sum_{t = 1}^T \log \frac{p_{\theta}(\ve{x}_{t-1}|\ve{x}_{t})}{q(\ve{x}_{t} | \ve{x}_{t-1})}\Bigg]} + \text{const}\\
     & \geq \mathbb{E}_q[-\log p_{\theta}(\ve{x}_{0})] + \text{const},
\end{split}    
\end{equation}
where Jensen’s inequality holds in the last line, and the $D_\text{KL}(\cdot||\cdot)$ is the \ac{kl} divergence, which computes the difference between two probability distributions. The objective of the optimization procedure is to minimize the loss function $L$ to reduce the difference between the probability distribution $q(\ve{x}_0)$ and the parameterized distribution $p_\theta(\ve{x}_0)$.

\subsection{Diffusion models in discrete state-spaces}\label{sec:dmdiscrete}

\begin{figure}
    \centering
    \includegraphics[width=\linewidth]{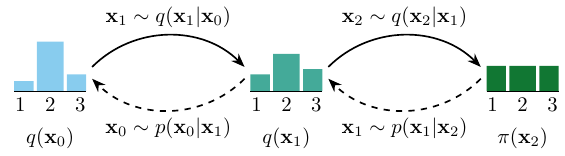}
    \caption{Example of \acp{dm} for discrete data. An initial data distribution $q(\ve{x}_0)$ is transformed into a uniform categorical distribution $\pi(\ve{x}_2)$ after $T=2$ time steps. The forward transition kernel is $q(\ve{x}_{t}|\ve{x}_{t-1})$, while $p(\ve{x}_{t-1}|\ve{x}_{t})$ is the transition kernel of the backward process obtained by training an \ac{ann}.}
    \label{fig:DM_discrete_state}
\end{figure}

\acp{dm} in \emph{discrete state-spaces} have been introduced by~\cite{Sohl2015} for binary random variables, and then generalized to $K$-categorical random variables~\citep{austin2021structured,hoogeboom2021} as depicted in~\cref{fig:DM_discrete_state}. 
In the general framework, given a scalar discrete $K$-category random variable $X$ taking values $x_{t} \in 1, \dots, K$ for $t \in [0, T]$, the forward transition probabilities from the category $i$ to the category $j$ at time $t$ can be realized by matrices:
\begin{equation}\label{eq: transition_probabilities_matrix}
    [\ve{Q}_{t}]_{i, j} = q(x_{t} = j|x_{t-1} = i).
\end{equation}
\change{Let $x_t$ be the scalar categorical variable at time $t$, we denote with $\mathbf{x}_t$ its one-hot encoded representation. Specifically, $\mathbf{x}_t$ is a row vector whose $k$-th component equals $1$ if $x_t = k$ and $0$ otherwise. The sample $\mathbf{x}_{t}$ at time $t$ is obtained by drawing from the forward transition kernel:}
\begin{align}
    \ve{x}_t &\sim  q(\ve{x}_{t}|\ve{x}_{t-1}), \\
    q(\ve{x}_{t}|\ve{x}_{t-1}) &= \text{Cat}(\ve{x}_{t}; \ve{p} = \ve{x}_{t-1}\ve{Q}_{t}),\label{eq:Cat_Kernel}
\end{align}
where Cat$(\ve{x}; \ve{p})$ is the categorical distribution sampling $\ve{x}$ with probability $\ve{p}$ and $\ve{x}_{t-1}$ is the sample at time $t-1$. 
Moreover, it can be shown~\citep{hoogeboom2021, austin2021structured} that a closed-form of the categorical \emph{posterior} can be computed:
\begin{equation}\label{eq: closed-form of categorical_postirior}
    q(\ve{x}_{t-1}|\ve{x}_t, \ve{x}_0) = \dfrac{q(\ve{x}_{t}|\ve{x}_{t-1}, \ve{x}_0) q(\ve{x}_{t-1}|\ve{x}_0)}{q(\ve{x}_t|\ve{x}_0)},
\end{equation}
where $q(\ve{x}_{t}|\ve{x}_{t-1}, \ve{x}_0) = q(\ve{x}_{t}|\ve{x}_{t-1})$ due to the Markov property and $q(\ve{x}_t|\ve{x}_0)=\text{Cat}(\ve{x}_{t}; \ve{p} = \ve{x}_{0}\ve{\Bar{Q}}_{t})$ with $\Bar{\ve{Q}}_t := \ve{Q}_{1}\ve{Q}_{2} \dots \ve{Q}_{t}$ the cumulative product of the transition matrices.

The denoising process is implemented via an \ac{ann} predicting the logits of the parameterized distribution:
\begin{equation}\label{eq:backwardProc}
    p_\theta(\ve{x}_{t-1}|\ve{x}_t),
\end{equation}
which has the functional form of a categorical distribution~\citep{hoogeboom2021, austin2021structured}.
The optimization procedure is realized by minimizing the loss function:
\begin{equation}
    L = D_\text{KL}(q(\ve{x}_{t-1}|\ve{x}_t, \ve{x}_0) ||p_\theta(\ve{x}_{t-1}|\ve{x}_t)).    
\end{equation}

\subsection{Classical Random Walk on Graph}
A graph is a pair $G=(V, E)$, where $V$ is a finite and non-empty set whose elements $v_i$, $i = 1, \dots, \abs{V}$, are called vertices (or nodes), and $E$ is a non-empty set of unordered pairs of vertices $e_{ij} = \{v_i, v_j\}$, called edges (or links). The \emph{order} and the \emph{size} of a graph are the cardinality $\abs{V}$ and $\abs{E}$ of the sets $V$ and $E$, respectively. The number $d_v$ of edges connected to the vertex $v$ is called \emph{degree} of the vertex. \change{The diameter of a connected graph $G$ is defined as $D = \max_{u,v \in V(G)} d(u,v)$, where $d(u,v)$ denotes the length of the shortest path between the vertices $u$ and $v$~\citep{west2001GraphTheoryIntroduction}.}
A graph is called \emph{undirected} if the edges of the graph do not have a direction, or \emph{directed} otherwise.  
A graph is completely defined by its \emph{adjacency matrix} $A$ that contains information on the topology of the graph, and whose element are defined as follow:
\begin{equation}
    [A]_{i, j} =\begin{cases}
    1, & \text{if} \quad e_{ij} \in E.\\
    0, & \text{if} \quad e_{ij} \not\in E.
  \end{cases}
\end{equation}

In a discrete-time \ac{crw} on an undirected graph $G$, at each step $t \in \mathbb{N}$ a walker jumps between two connected nodes $v_i, v_j$ with some probability that is described by the stochastic transition matrix $S = [S]_{i,j}$~\citep{Weiss1982, GeorgeHerbertWeiss1994, Caruso2014}. The matrix $S$ is related to the adjacency matrix by the equation: $[S]_{i,j} = [A]_{i,j}/d_i$, \change{ where $d_i$ denotes the degree of vertex $v_i$, i.e. the number of edges incident to $v_i$, and it is given by $d_i = \sum_{k=1}^{|V|} [A]_{i,k}$}.
Formally, a walker is represented by a discrete random variable $X$, and its trajectory after the $T$ time steps is the set $\{x_1, \dots , x_T\}$, where the value $x_t$, with $t=1, \dots, T$, corresponds to the node occupied by the walker at time $t$.
A \ac{crw} is a Markov process, that is, the distribution at time $t+1$ depends only on the distribution at time $t$.
Given $\ve{p}_{t}=\{p_i^{(t)}\}$, the occupation probability distribution of a walker over the nodes $v_i$ after $t$ time steps, the evolution of the distribution at time $t+1$ is given by:
\begin{equation}\label{eq: CRW_eq}
    \ve{p}_{t+1}= S \ve{p}_{t}.
\end{equation}
The distribution of a \ac{crw} converges to a limiting \emph{stationary} solution:
\begin{equation}\label{eq: CRW_stationary_solution}
    \pi =  \Bigg\{\dfrac{d_i}{2\abs{V}} \Bigg \}, \quad t\rightarrow \infty,
\end{equation}
independently of the initial distribution $\ve{p}_0$.
For a $d-$regular graph (all vertices have the same degree \change{$d_i=d$ with $i=1,\dots,|V|$}) the limiting stationary distribution is \emph{uniform} over the nodes of the graph.

\subsection{Quantum Walks}

In quantum information theory, \aclp{qw} were introduced by~\cite{Aharonov1993qw} as quantum analogues of classical random walks. However, in \acp{crw} the dynamics is purely stochastic at each time step, while \acp{qw} evolve via a deterministic dynamics, and the stochasticity comes out only when a measurement is performed on the quantum state of the walker~\citep{Aharonov1993qw, Kempe2009}. Moreover, \acp{qw} involve peculiar quantum mechanical properties such as \emph{coherent superposition}, \emph{entanglement}, and \emph{quantum interference}, resulting in a faster spread (ballistic for the quantum case, while diffusive for the classical one).
There exist two different formulations of \acp{qw}: i) \acp{ctqw} and ii) \acp{dtqw}. In the former, the unitary evolution operator can be applied at any time $t$, while in the latter, the operator can be applied only in discrete time steps. Furthermore, \acp{dtqw} need an extra degree of freedom, called ``coin'', which stores directions and speeds up the dynamics of the walker~\citep{Ambainis2005}.

\subsubsection{Discrete-Time Quantum Walks on Graph}

\begin{figure}[h!]
    \includegraphics[width=\linewidth]{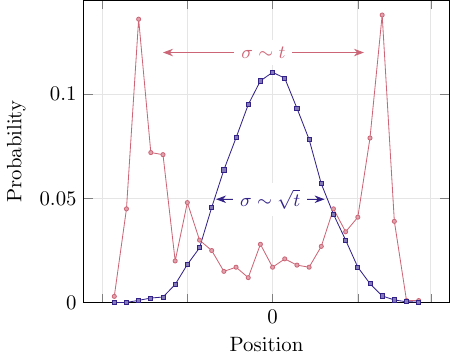}
    \caption{Comparison between the Gaussian probability distribution (violet squares) obtained from a \ac{crw} and probability distribution of a \ac{qw} with Hadamard coin (red circles).}
    \label{fig:gaussian_vs_quantum}
\end{figure}

Given a graph $G(V,E)$, let $\mathcal{H}_{V}$ be the Hilbert space spanned by the position states $\{\ket{v} : v = 1, \dots, \abs{V}\}$, and let $\mathcal{H}_A$ be an auxiliary Hilbert space spanned by the coin states $\{ \ket{\downarrow}, \ket{\uparrow} \}$.
The total Hilbert space $\mathcal{H}$ associated to a \ac{qw} is obtained by the tensor product between the auxiliary space and the position space:
\begin{equation}
    \mathcal{H} = \mathcal{H}_A \otimes \mathcal{H}_V.
\end{equation}
In general, a state is written as:
\begin{equation}
    \ket{\psi} = \ket{a} \otimes \ket{v}, \quad \ket{a} \in \mathcal{H}_A \quad \ket{v} \in \mathcal{H}_V.
\end{equation}
The dynamics of the quantum walker are governed by the unitary single time-step operator $\hat{U}$ acting on the total Hilbert space:
\begin{equation}
     \hat{U} = \hat{S} \cdot (\hat{C} \otimes \hat{I}),
\end{equation}
where $\hat{I}$ is the identity on position space, $\hat{C}$ is the coin operator acting on the auxiliary space, and $\hat{S}$ is the shift operator acting only on position space and moving the walker from state $\ket{v}$ to state $\ket{v+1}$, or $\ket{v-1}$. Formally, the shift operator is given by:
\begin{equation}
    \hat{S}=\sum_v \ketbra{\uparrow}{\uparrow }\otimes \ketbra{v+1}{v}+\ketbra{\downarrow }{\downarrow } \otimes \ketbra{v-1}{v}.
\end{equation} 
The coin operator $\hat{C}$ can be chosen from the family of unitary operators, and its choice leads to symmetric (unbiased walk) or asymmetric (biased walk) distributions. A common choice for $\hat{C}$ is the Hadamard coin:
\begin{equation}
    \hat{H} = \frac{1}{2}\left( \begin{array}{cc} 1 & 1 \\
     1 & -1 \end{array} \right),
\end{equation} which leads to an unbiased walk.
The state of a quantum walk after $t$ discrete time steps is given by:
\begin{equation}\label{eq:evolution_qws}
    \ket{\psi(t)} = \hat{U}^{\otimes t} \ket{\psi(0)}.
\end{equation}
In general, \acp{qw} do not converge to any limiting distribution, in contrast to the classical ones:
the limit $\lim_{t\to\infty}\ket{\psi(t)}$ does not exist due to the unitary evolution~\citep{aharonov2001quantumwalkgraph}. Moreover, the interference effects lead a quantum walker to spread \emph{quadratically} faster with respect to its classical counterpart. Namely, in the classical case, after $t$ time steps, the expected distance from the origin is of order $\sigma \sim \sqrt{t}$, while in the quantum case it is of order $\sigma \sim t$, as shown in \cref{fig:gaussian_vs_quantum}

\subsection{Quantum Stochastic Walks on Graph}\label{sec:qsw} 
\acp{qsw} are introduced by~\cite{whitfield2010quantum} and describe the evolution of a walker following quantum stochastic dynamics. The class of \acp{qsw} generalizes and encompasses both \acp{crw} and \acp{qw}. 

Formally, a continuous-time \ac{qsw} dynamics is described by the Kossakowski–Lindblad-Gorini master equation~\citep{KOSSAKOWSKI1972247, Lindblad1976, Gorini1976}:
\begin{equation}\label{eq:qsw}
    \frac{d\rho}{dt} = \change{-}(1-\omega)i[H, \rho] + \omega \sum_j L_j\rho L_j^{\dagger} -\frac{1}{2} \{L_jL_j^{\dagger},\rho\},
\end{equation}
where $\rho$ is the walker density matrix, $H$ is the Hamiltonian of the system describing the coherent evolution, and $L_j$ are Lindblad operators responsible for the incoherent dynamics, which represent the interactions of the system with an external environment. The continuous parameter $\omega\in [0, 1]$ quantifies the interaction between coherent and incoherent evolution. 
For $\omega = 0$, \cref{eq:qsw} describes the evolution of a pure \ac{qw}. Instead, setting $\omega = 1$ and choosing $L_{ij} = S_{ij} \ketbra{i}{j}$, \cref{eq:qsw} describes the motion of a \ac{crw}, where $\ket{i}$ is the quantum basis state associated with node $i$ of a graph and $S_{ij}$ is the transition matrix of the walker from node $j$ to $i$. 

\change{To summarize, \acp{qsw} represent a useful model to compare quantum, hybrid and classical dynamics under similar setups by tuning only one parameter.}

\section{Results}\label{sec:results}

\subsection{Quantum, hybrid and classical stochastic diffusion}\label{sec:resultsQSW}
In this section, we study the performance of the image generation task by \acp{dm} when the classical stochastic dynamics in the forward chain are replaced by, or interplay with, quantum stochastic diffusion processes. For this purpose, we decide to adopt the formalism of \acp{qsw} of \cref{sec:qsw}, which provides a useful tool to study the transition from classical to quantum diffusion dynamics, and the categorical \acp{dm} of \cref{sec:dmdiscrete}. This decision is inspired by previous works where the \ac{qsw} formalism is used to find an optimal mixing of classical and quantum dynamics for information transport~\citep{Caruso2014,Caruso2016,DallaPozza2022}. 

In the following, a categorical data sample is represented by a quantum stochastic walker that evolves on a cycle graph of $8$ nodes, which represent the categories of the sample. In this approach, the density operator $\rho$ of \cref{eq:qsw} describes the state evolution of the walker on the graph: the diagonal elements $\rho_{ii}$ give the probabilities that the walker is at the node $i$, while the off-diagonal elements $\rho_{ij}$ describe and contain information on the peculiar quantum mechanical effects (coherence) during the diffusion process. Next, we refer to the diagonal and off-diagonal elements of the density matrix $\rho$, respectively, as \emph{populations} and \emph{coherences}.

\begin{figure}[htp!]
    \includegraphics[width=\linewidth]{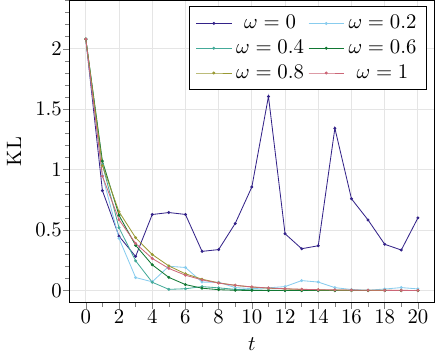}
    \caption{The \ac{kl} divergence between the populations of a single \ac{qsw} and the uniform distribution on a cycle graph of $8$ nodes for different value of $\omega$ after $T = 20$ time steps. The walker is initially in the node $0$, corresponding to the state $\rho = \ketbra{0}{0}$, and its evolution over the graph is obtained by solving the equation \cref{eq:qsw}.}
    \label{fig:klSingleWalker}
\end{figure}
In \cref{fig:klSingleWalker} we show the evolution of the \ac{kl} divergence between the populations of the quantum stochastic walker on a cycle graph of $8$ nodes and the corresponding uniform distribution for different values of $\omega$. 
In particular, we can observe that for pure quantum diffusion dynamics ($\omega=0$) the \ac{kl} divergence manifests an oscillating behavior, while in the classical case $(\omega = 1)$ it smoothly converges to zero. In hybrid scenarios, with $\omega \in (0, 1)$, the presence of the incoherent term of \cref{eq:qsw} dampens the oscillations of the pure quantum case, leading to faster convergence with respect to the classical case, for example, for $\omega = 0.4, 0.6$.

\subsection{Image generation}\label{sec:qswImg}
\begin{figure}[htp!]
    \centering
    \includegraphics[width=0.8\linewidth]{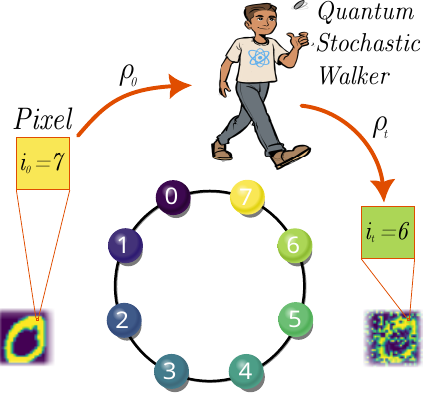}
    \caption{Illustration of the model. Each pixel of an image sample represents an independent quantum stochastic walker moving on a cycle graph of $8$ nodes that correspond to the gray intensity values. The walker moves on the graph by \cref{eq:qsw}.}
    \label{fig:pxl-qwalker}
\end{figure}

Starting from the results of the previous section, a natural question arises: How do the different \acp{qsw} dynamics impact the generation of new samples? To answer this question, we choose to perform image generation on the MNIST dataset, scaling the grayscale levels of pixels from $[0, 255]$ to $[0, 7]$. We propose a forward dynamics, illustrated in \cref{fig:pxl-qwalker}, where each pixel of the image is an independent \ac{qsw} on a cycle graph of $8$ nodes (one for each gray intensity value of the pixel).
In the following, for simplicity we describe the forward process for a single \ac{qsw}. 

The walker is initially at the node $i_0$, where $i_0 = 0, 1, \dots, 7$, and is described by the quantum state $\rho_0 = \ketbra{i_0}{i_0}$. Subsequently, the state of the walker evolves by \cref{eq:qsw}, and we collect the populations of the states $\rho_1,\dots,\rho_T$ that we use to define the forward categorical distributions, from which we sample the positions of the walker at each time step:
\begin{align}
    i_t &\sim \text{Cat}(\ve{p} = \text{diag}(\rho_{t-1})),\\
    \rho_t &= \ketbra{i_t}{i_t},
\end{align}
where $i_t$ is the walker position on the graph at time $t$, and $\rho_t$ is the corresponding quantum state.

The backward process is implemented \change{classically} with \ac{mlp}. \change{This choice is driven by the simple structure of the data samples and by our main goal, which is to investigate the impact of different hybrid diffusion dynamics, rather than the role of classical or quantum neural network architectures in the denoising process. For this reason, we opt for a minimal classical backward network and for an unconditioned generative setting. In the backward process, t}he \ac{mlp} takes as input the one-hot encoding of the walker position at time $t$ for all pixels of the image and is trained to predict the position at time $t-1$ for them \change{(a more detailed description is provided in Sec.~\ref{sec:ImplementationDetails})}.

\begin{figure}[t!]
    \centering
    \includegraphics[width=\linewidth]{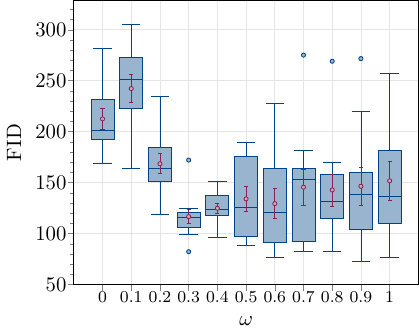}
    \caption{Box plot of \ac{fid} values for different values of $\omega$.
     Every box is obtained from $10$ different repetitions of for the same value of $\omega$ with $T=20$, $8$-cycle graph. Mean and standard error of the mean are also reported. The plot show how the hybrid quantum-classical diffusion dynamics $(\omega = 0.3)$ results to generate statistically better image datasets.}
    \label{fig:Fidboxplot}
\end{figure}

In order to evaluate the generation performance of different quantum stochastic dynamics, we compute the \ac{fid} metric~\citep{Szegedy2016FID} that assesses the quality of images created by a generative model. More precisely, the \ac{fid} metric calculates the distance between the original and the generated datasets, and is given by:
\begin{equation}
    \text{FID} = || \mu - \mu' ||^2_2 + \tr(\Sigma + \Sigma' - 2(\Sigma\Sigma')^{\frac{1}{2}}),
\end{equation}
where $\mu$ and $\mu'$ are, respectively, the mean of the multivariate normal distribution of the features of the original and generated image dataset, with $\Sigma$ and $ \Sigma'$ the corresponding variances. A higher value of \ac{fid} indicates poorer performance of the generative model. Moreover, to statistically assess the performance of different models, we implement $10$ simulations for each value of $\omega \in \{0, 0.1, 0.2, \dots, 1 \}$, and report in \cref{fig:Fidboxplot} the box plot visualization of the \ac{fid} values of the generated digits $0$ (box plots require at least $5$ samples to provide solid statistical information~\citep{Krzywinski2014boxplot}).

We observe that for a hybrid quantum-classical stochastic dynamics ($\omega=0.3$), the mean value of the \ac{fid} is lower than in the classical case ($\omega=1$). Moreover, the box plot shows that in the hybrid scenario the \ac{fid} values of the $10$ simulations are closely distributed around the median, while in the classical case most of the \ac{fid} values are above it. In addition, all simulations for $\omega=0.3$, except for the single upper outlier, result in better \acp{fid} than half of the runs for $\omega=1$. This means that the model with a specific interplay of quantum-classical dynamics is able to steadily generate better samples and is also statistically more robust than the classical one. 

\change{The better performance in the image generation of the hybrid dynamics for $\omega = 0.3$ can be interpreted by noting that a balance between quantum coherence and stochastic noise enables a more efficient information transport across the graph. Furthermore, the value found is consistent with the values of $\omega$ reported in~\cite{Caruso2014, Caruso2016, DallaPozza2022}. Also, it is shown in~\cite{Caruso2014} that for a large class of graphs (including random graphs, small‑worlds, rings, stars, etc.) with diameter $D > 3$, the transport efficiency of a \ac{qsw} reaches its highest value typically around $\omega \sim 0.1$, whereas for pure classical dynamics ($\omega = 1$) it is optimal for fully‑connected graphs $D = 1$. In our case, the eight‑node cycle graph considered here satisfies the condition $D = 4 > 3$, and therefore the better performance of the hybrid dynamics with $\omega = 0.3$ observed in \cref{fig:Fidboxplot} can be explained within this framework. In other words, more efficient transport results in better diffusion over the graph, allowing for a more complete exploration of the state space, leading to better representations.}

\begin{figure*}[htp!]
    \centering
    \includegraphics[width=\linewidth]{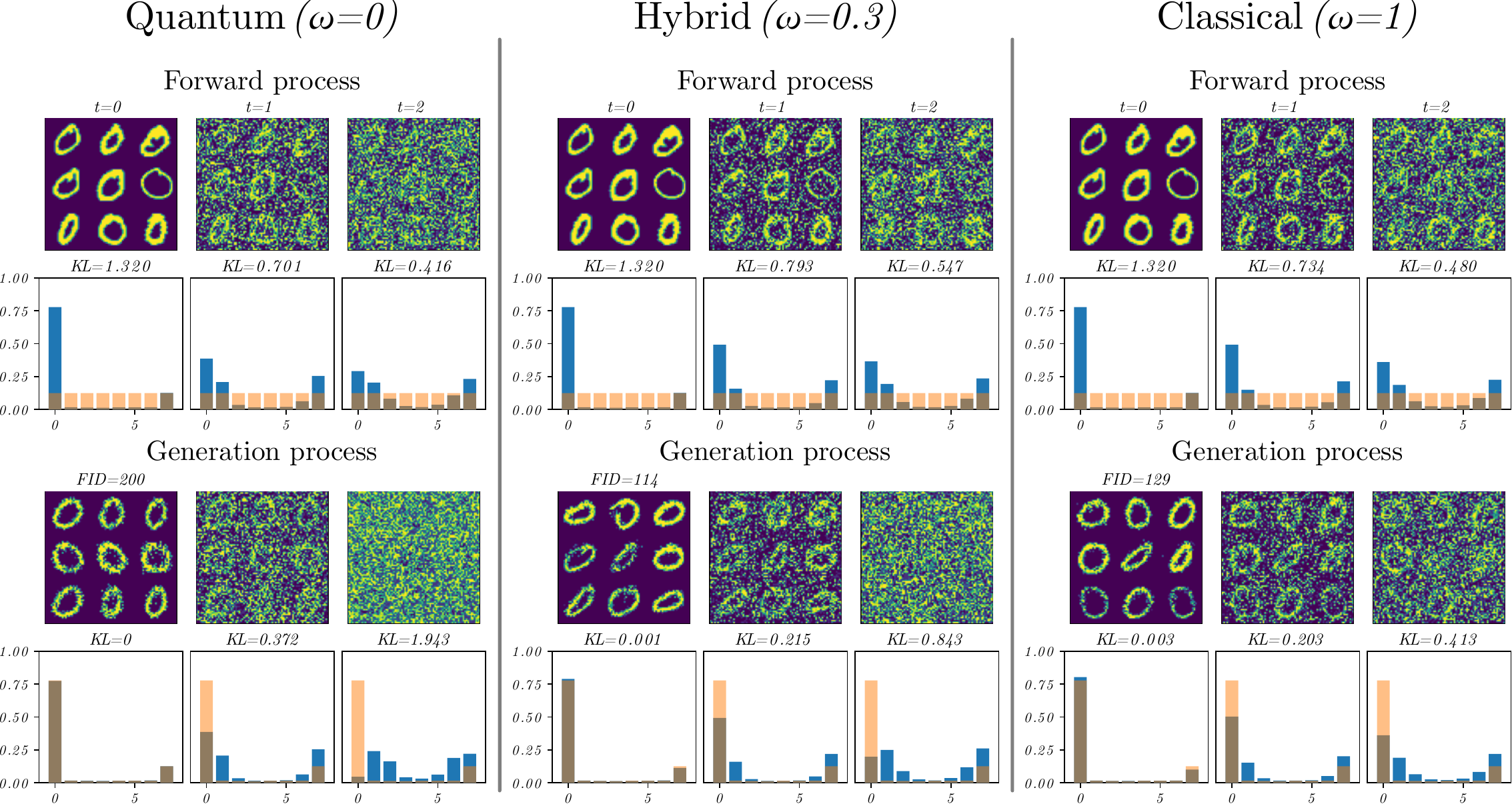}
    \caption{Image generation with \ac{qsw}-based \acp{dm} via a quantum ($\omega =0$), hybrid $(\omega = 0.3)$ and a classical ($\omega =1$) forward chain of $T=20$ time steps. We use the models with the inferior median values of \ac{fid} between the 10 repetitions of \cref{fig:Fidboxplot}. We report the first two steps of the forward chain (in the first row) of 9 different samples, and the evolution of the distribution of the pixel values for the entire dataset of digits $0$ of MNIST (blue bins in the second row) that is compared with the final uniform prior (orange bins in the second row). The \ac{kl} divergence value between the two distributions is reported on the top. We also illustrate the final two steps of  9 generated samples, and the \ac{fid} score between the entire training dataset and generated dataset is reported (third row). In the last row, we compare the distribution of the pixel values of the entire generated dataset (blue) with the training dataset distribution (orange), and \ac{kl} divergence between them.}
    \label{fig:qswEvolution}
\end{figure*}

Finally, in \cref{fig:qswEvolution} we show $9$ random samples of generated images of the digit $0$ using quantum, hybrid, and classical stochastic dynamics.

\subsection{Implementation on NISQ Devices}\label{sec:resultsQW_NISQ}
A hybrid \ac{qsw} dynamics can be interpreted as a \ac{qw} interacting with an external environment introducing noise. In this section, we therefore perform image generation by implementing a \ac{qw} that exploits the intrinsic noise of a \ac{nisq} device in the forward process. This dynamic is efficiently implemented on a cycle graph with the quantum circuit of~\cite{Razzoli2024}.
The efficiency of the algorithm allows us to modulate the amount of noise introduced in the forward chain, which has low values by default and is increased by delays. As in \cref{sec:qswImg}, we represent each pixel by a \ac{qw} moving on a cycle graph of $8$ nodes. Furthermore, the rotation-invariant property of the cycle graph allows us to run each walker as if starting from the same initial state, and then re-mapping the outcome to the specific value of the pixel color by a shift operation. In this way, the forward chain is run only once for all the \acp{qw}, a condition that is necessary to achieve our algorithm on the limited availability of the current quantum processors. 
More precisely, our model requires only $4$ qubits to generate $28 \times 28$ grayscale MNIST images (normalized to $8$ gray values): $3$ qubits for the position of the \ac{qw} on the $8$ nodes of the graph and $1$ qubit for the coin's degree of freedom of the quantum walker. The mapping of the pixel intensity values into the positions on the graph allows us to introduce quantum effects in the pixel dynamics, as well as to make our model \change{\emph{scalable}} in image size with respect to other \acp{qdm} approaches for generating images~\citep{kölle2024quantumdenoisingdiffusionmodels, cacioppo2023quantumdiffusionmodels}. 
\begin{figure}[ht!]
    \includegraphics[width=\linewidth]{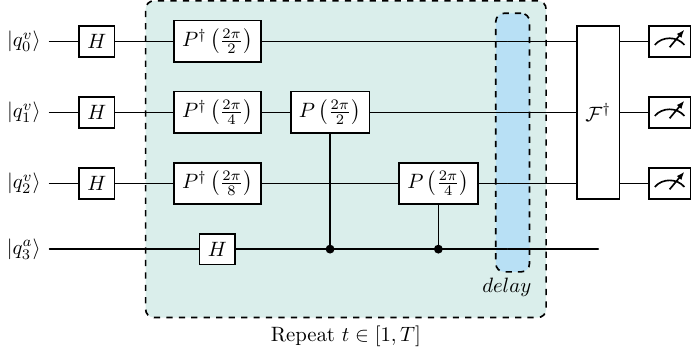}
    \caption{Implementation of discrete time \ac{qw} on cycle graph using quantum circuits. The $H$ and $P$ are respectively Hadamard and phase gates, and $\mathcal{F}$ is the quantum Fourier transform. The outer shaded block indicates a single time step of the walker and is repeated for $t$ times up to $T$. The inner shaded block modulates the amount of noise in the circuit by a delay operation scaled as in \cref{eq:delay}. The final measurements are also shown.}
    \label{fig:circuit}
\end{figure}
In \cref{fig:circuit} we depict the quantum circuit used for the implementation of the \ac{qw}. At the end of the circuit, we collect measurements on the position qubits to obtain the distribution of the walker at every $t \in [1, T]$. The latter is used to define the categorical function from which we draw the new positions of the walker on the graph.

The noise in the circuit is injected with \emph{delay} operations expressed in seconds as follows:
\begin{align}
    delay &= c \times scaling\times dt,\label{eq:delay}\\
    scaling &= \left\lfloor\frac{\sin^2(\frac{\pi}{2} \frac{t}{T-1})}{8}\right\rfloor 8,
\end{align}
where the value of $scaling$ is truncated to the nearest integer multiple of 8 for hardware reasons, and $dt=5\cdot 10^{-10}$ is the time length in seconds of a single operation on the device used. The value of $c$ is a coefficient used to ensure the convergence of the forward process to the uniform distribution within a previously fixed number of time steps $T=20$, and we choose $c=5 \cdot 10^4$. This scaling of the injected noise is chosen in analogy to the cosine schedule of noise in the classical \ac{dm} of~\cite{nichol21}.
The backward process of the \ac{qw}-based diffusion model is implemented with a \ac{mlp} analogous to the one used for the \ac{qsw}-based case of \cref{sec:qswImg}.

As a proof of concept, we train our model with $6\,903$ full-size $28 \times 28$ MNIST images of digits $0$. The forward process is implemented in \emph{Qiskit}~\citep{qiskit2024}, and is initially simulated on \emph{fake\_brisbane} with $10^5$ shots, and then run on the real device \emph{ibm\_brisbane} with $10^4$ shots. The number of shots is chosen as the maximum available for the simulator and subsequently reduced considering the computational resources available for executing the algorithm on the \ac{nisq} hardware. This forward protocol is well suited for the \ac{qpu} used because it has a maximum connectivity degree of 3 and the coin qubit can interact directly with all position qubits. As an additional advantage, it is also possible to run several walkers in parallel on different parts of the same \ac{qpu}. 

\begin{figure*}[ht!]
    \centering
    \includegraphics[width=\linewidth]{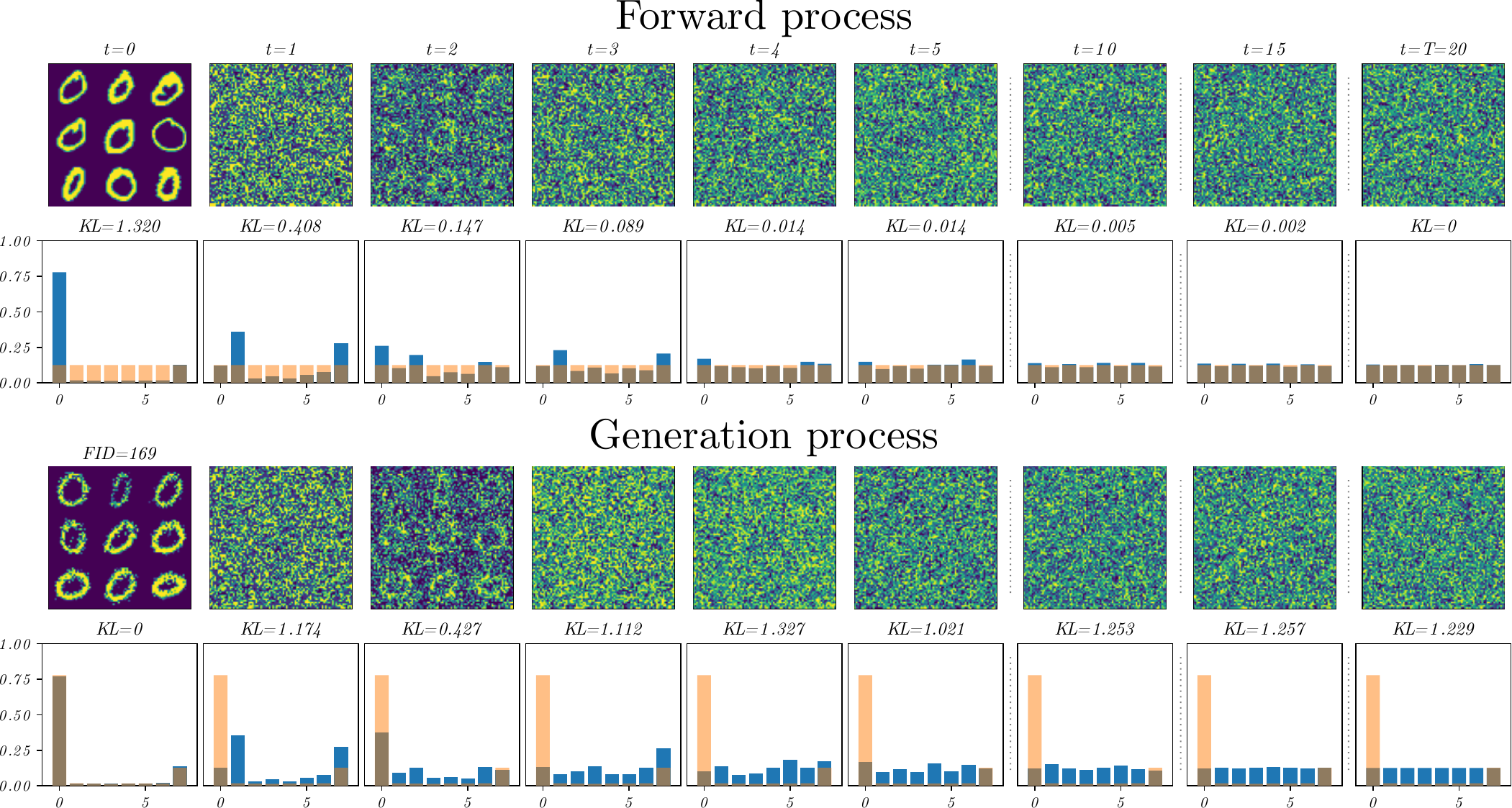}
    \caption{Images generation with \ac{qw}-based \ac{dm} with noise from real \emph{ibm\_brisbane} \ac{nisq} device.
    We report for selected values of $t$ the evolution of 9 random samples of the real dataset in forward (first row) and 9 random generated samples in backward (third row). In the second row we show the evolution of the pixel values distributions (orange) for the entire training dataset reporting the \ac{kl} divergence with the uniform distribution (blue). Finally, in the fourth row we show the value distributions of a generated dataset of the same dimension of the training set comparing it with the original dataset using the \ac{kl} divergence. Also, we report for $t=0$ the value of \ac{fid} equal to \change{$169$} between the original and generated dataset.}
    \label{fig:digitEvolution}
\end{figure*} 
In \cref{fig:digitEvolution} we show the forward and backward processes for image generation with diffusion implemented on the real IBM machine. The equivalent figure obtained using the simulator is available in \cref{sec:appendix} together with the results on the other digits of MNIST. In the first and third rows show, respectively, the quantum forward evolution via \acp{qw} (from left to right) and the image generation by a trained classical \ac{mlp} (from right to left) at different time steps $t$ for 9 samples of the original and generated dataset. In the second row, the transformation of the initial distribution of the pixel values (in blue) for the full training dataset overlaps with the desired final uniform distribution (in orange) with the \ac{kl} divergence between the two reported on the top. The final row reports the distribution of the pixel values for $6\,903$ generated images (in blue) overlapped with the original distribution of the training dataset (in orange) with the \ac{kl} divergence between the two on top. In the top distribution at $t=0$ we report the value of \ac{fid}$ = 169$ calculated between the original dataset and the equally sized dataset of generated images.

Comparing the results of \cref{fig:qswEvolution} with \cref{fig:digitEvolution}, we observe that the latter generates samples with higher values of \ac{fid}: $114$ (hybrid \ac{qsw}-based) and \change{$169$} (IBM-based \ac{qw}). One possible explanation for the difference between the simulated results and the real IBM-based ones could be that the underlying dynamics for the evolution of the pixel values is in discrete time for the \acp{qw} which are implemented with circuits, while it is in continuous time for \acp{qsw}, resulting in a more gradual introduction of noise during the forward process.

\section{\change{Implementation Details}}\label{sec:ImplementationDetails}
\begin{figure}
    \centering 
    \includegraphics[width=\linewidth]{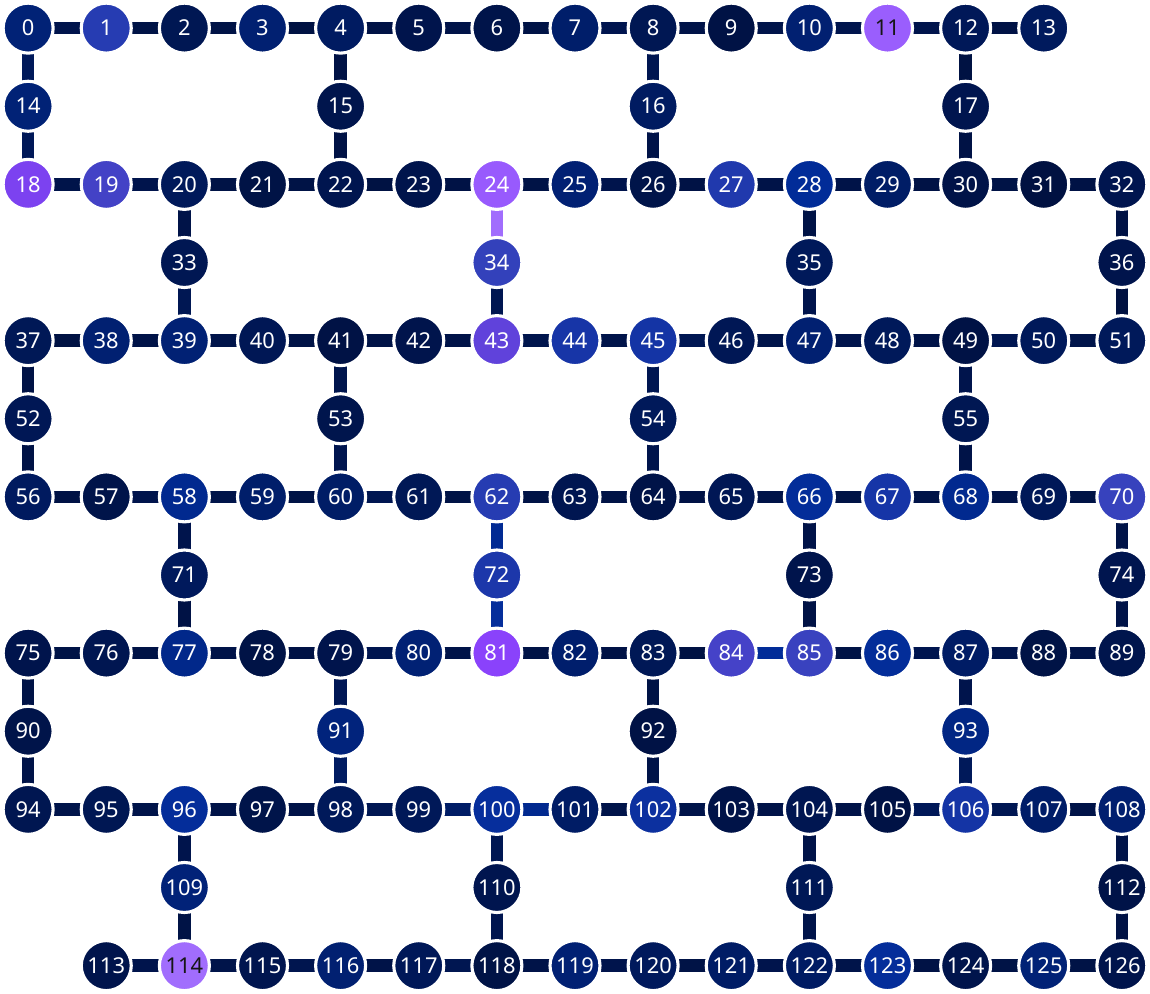}
    \caption{Qubit connectivity map of the real \ac{qpu} \emph{ibm\_brisbane} (courtesy of IBM~\citep{ibmq}). Every circle represents a qubit, and lines represent their connections. 
    Colors code the readout errors (circles) and the errors for the connections (lines). Dark blue indicates a small error, while large one is in purple.}
    \label{fig:ibm_brisbane_QPU}
\end{figure}
The implementation is carried out in Python 3 using Qiskit~\citep{IBM_Qiskit}, an IBM open-source software to work with real quantum processors at the circuit level, QuTiP~\citep{QuTip}, an open-source computational physics software to simulate open quantum systems dynamics, and PyTorch~\citep{PyTorch}, which is a flexible and efficient machine and deep learning library. 

\change{\subsection{Quantum forward process}}
For the implementations of \acp{dm} via \acp{qsw} on a cycle graph, we used the QuTip function \emph{mesolve} to compute the evolution of the state $\rho_t$ of the quantum stochastic walker in \cref{eq:qsw} for different values of $\omega$ and at consecutive time steps, fixing $\delta t =6 \cdot 10^{-1}$.
Regarding the forward process with \acp{qw} in \cref{sec:resultsQW_NISQ}, we use the Qiskit library to implement the \acp{qw} dynamics both simulated and on the real IBM \ac{qpu} \emph{ibm\_brisbane}, whose hardware topology is shown in \cref{fig:ibm_brisbane_QPU}.

\change{\subsection{Classical backward process}}
The backward process is implemented via \acp{mlp} of PyTorch linear layers with \ac{relu} activation functions. More precisely, the architecture is structured with a head layer of size $800$ shared between all time steps $t$ and two tail layers of the same size \change{$800$,} specialized for each  step. The input layer takes the one-hot encoding of all the positions of the walkers at the time step $t$ for all the image pixels, and the final layer predicts their logits at the previous time $t-1$.
\change{In detail, the forward works at single-pixel level, with independent walkers associated to each pixel. Instead, the backward works at image level. The \ac{mlp} takes as input a batch of full $28\times 28$ images at time $t$ obtained from the forward, and predict a batch of all $28\times 28\times 8$ logits for the possible intensity levels of all the pixels of the image at time $t-1$.}
\change{In the training, we want that the \ac{mlp} predicts with high probability the pixels intensity level of the images obtained from the forward process. To achieve this, we minimize the categorical cross-entropy loss} for $10^4$ epochs using \emph{Adam}~\citep{kingma2017adam} with a batch size of 16 samples and setting the learning rate equal to $10^{-3}$.

\section{Discussion and Conclusions}
In this work, we show that the impact of quantum noise within the forward \acp{dm} dynamics influences the effectiveness of the backward generation process. In particular, we find that a model with a hybrid \ac{qsw} diffusion dynamics (\change{$\omega=0.3$}) produces sets of samples with a lower \ac{fid} and is also statistically more solid than its classical counterpart. \change{This result can be explained by the fact that a balance between quantum coherence and stochastic noise allows for more efficient information transport across the graph, producing better image sets for diffusion models. Furthermore, the value is consistent with the results of~\cite{Caruso2014, Caruso2016, DallaPozza2022} where it is observed that efficient information transport is achieved by means of hybrid \acp{qsw} dynamics.}

Successively, we propose a model that allows one to generate images of any size by harnessing the noise of real \ac{nisq} devices, taking into account the topology and connectivity of the \ac{qpu}. More precisely, \change{our quantum algorithm uses $q+1$ qubits for generating images with intensity values in range $[0,2^q-1]$ where $q \in \mathbb{N}^+$. Therefore, the number of qubits is related only to the number of grayscale pixel levels, and not to the image dimensions. In our study,}  we generate $28 \times 28$ gray-scale MNIST images using only $4$ qubits by efficiently implementing a \ac{qw} dynamics on a cycle graph whose nodes are the color values of a single image pixel. The invariant property of the graph allows one to independently run the walkers for the single pixels starting from the same initial state and then remapping the outcome to the specific value of the pixel color. Furthermore, our forward protocol enables the implementation of multiple walkers in parallel, requiring a maximum $3$ degree of connectivity between qubits of a \ac{qpu}. This allows for the implementation on the currently available \ac{nisq} devices \change{and} we show how noise can be used as a resource in the context of \ac{q-gen-ai} and not only be a detrimental factor for quantum algorithms.
\change{In addition, we believe that our hybrid strategy can be useful in order to adopt already existing noisy quantum processors while simultaneously avoiding the common pitfalls of quantum learning models such as barren plateau, robustness and trainability.}

Some future research directions can focus on a major integration of the \ac{qpu} topology with the \ac{qw} implemented at the circuit level. In particular, we can take more advantages from the connectivity to increase the range of the pixel values, and thus generate high-quality images. Moreover, error correction or mitigation techniques could be used to better control the level of noise in the quantum forward to further improve the capabilities of the backward network. \change{We have evidence of the importance of noise characterization by comparing the FID values for images generated using the real IBM quantum hardware \emph{ibm\_brisbane} and the simulator \emph{fake\_brisbane}, as shown in \cref{fig:digitEvolution} and \cref{fig:digitEvolution_sim,fig:digitEvolution_sim_1_9-samples} in \cref{sec:appendix}.}
Another interesting outlook can be the possibility of a physical realization of the quantum walk, making our model directly applicable to quantum data without performing any quantum embedding in the first stage of the algorithms, and then using a quantum \acp{ann} in the reverse process. In conclusion, we believe that the latter can be fruitful where it is necessary to learn unknown quantum phenomena, for instance, in quantum sensing, metrology, chemistry, and biology scenarios.

\change{During the revision process, we further developed, in collaboration with another research group, the \ac{qw}-based diffusion model presented in this work. In particular, we incorporate new methodological contributions and an evaluation based on more structured medical imaging datasets~\citep{Venturelli2026}.}

\appendix
\section{\change{Appendix}}\label{sec:appendix}
\begin{figure}[h!]
    \centering
    \includegraphics[width=\linewidth]{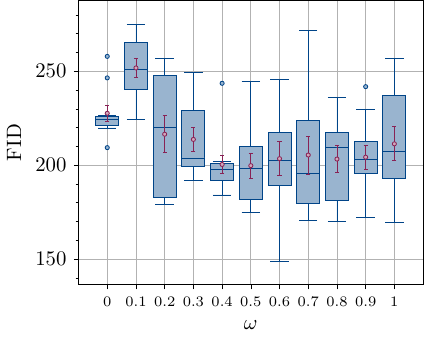}
    \caption{\change{Box plot, mean and standard error of the mean of the \ac{fid} values for $\omega=0,0.1,\dots,1$ obtained with the same setting used in \cref{fig:Fidboxplot}, but for the entire MNIST dataset and with a model trained for 100 epochs.}}
    \label{fig:FidboxplotFullMnist}
\end{figure}

\change{Here we report further obtained results. In detail, in \cref{fig:FidboxplotFullMnist} we show the performances of diffusion models with \ac{qsw} forward with different values of $\omega$ and trained on the full MNIST dataset (all digits included). Also in this case, as in \cref{fig:Fidboxplot}, we observe that for a particular hybrid dynamics ($\omega=0.4$) results in statistically more robust models that generate images with better \ac{fid} values. Due to the increase in computational resources (we use all the digits instead of only the zero), the models are trained for $100$ epochs instead of the $10^4$ used in \cref{fig:Fidboxplot}. This results in higher \ac{fid} values comparing \cref{fig:FidboxplotFullMnist} and \cref{fig:Fidboxplot}.}

\begin{figure}[h!]
    \centering
    \includegraphics[width=\linewidth]{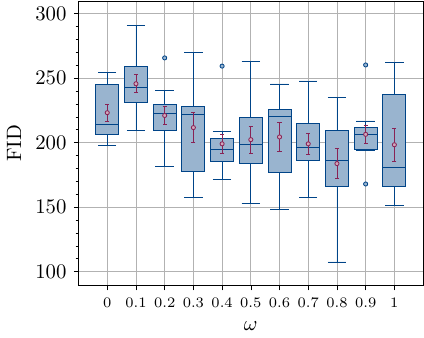}
    \caption{\change{Box plot, mean and standard error of the mean of the \ac{fid} values for $\omega=0,0.1,\dots,1$ obtained with the same setting used in \cref{fig:Fidboxplot}, but for the EMNIST-letters dataset and with a model trained for 100 epochs.}}
    \label{fig:FidboxplotEMnist}
\end{figure}

\change{To have a broader assessment of the applicability of our approach to different datasets, we trained also the model on the EMNIST-letters dataset~\citep{Cohen2017} which is composed of handwritten letters instead of digits. The result of the training for $100$ epochs is visible in~\cref{fig:FidboxplotEMnist}. The same trend observed in~\cref{fig:Fidboxplot,fig:FidboxplotFullMnist} is still visible with this dataset, although the advantage of the hybrid model with respect to the classical one is slightly less prominent.}

\begin{figure*}[h!]
    \centering
    \includegraphics[width=\linewidth]{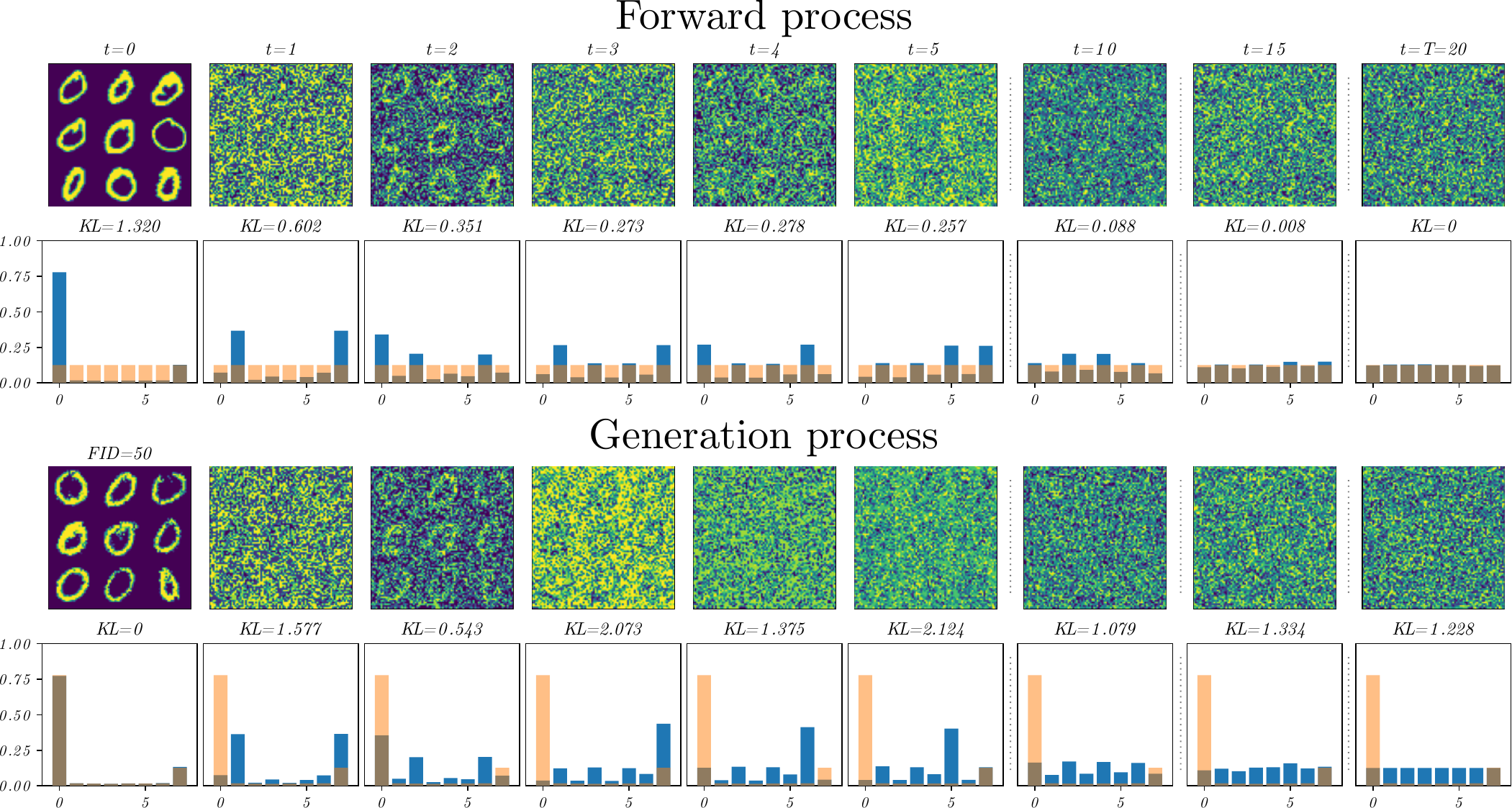}
    \caption{\change{Images generation with \ac{qw}-based \ac{dm} with noise from IBM simulator \emph{fake\_brisbane}.}
}
    \label{fig:digitEvolution_sim}
\end{figure*}

\begin{figure*}[h!]
    \centering
        \includegraphics[width=0.94\linewidth]{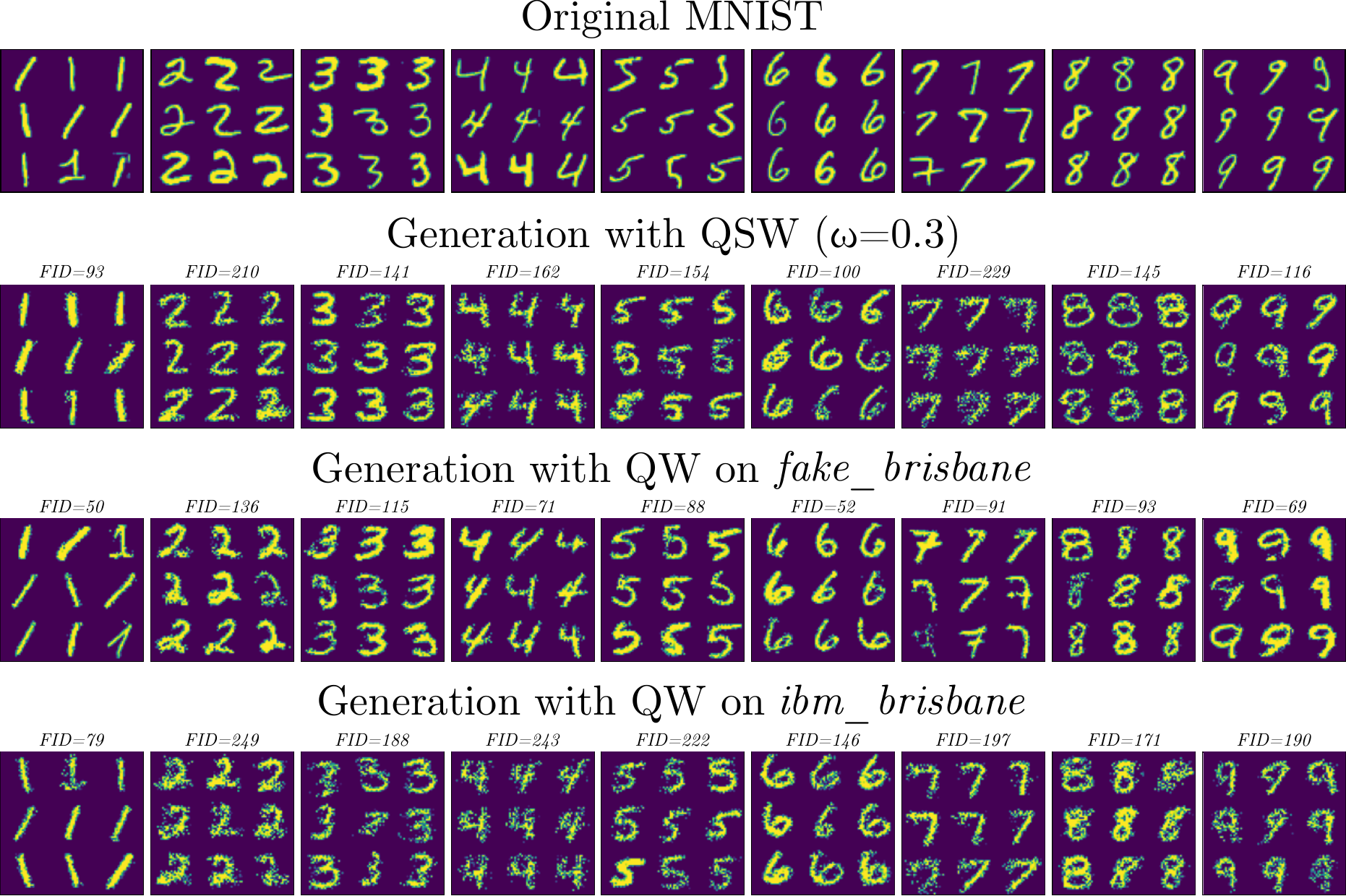}
    \caption{\change{Original MNIST images (first row) and images generated with a hybrid \ac{qsw} dynamics (second row) and a \ac{qw}-based \ac{dm} with noise from a simulated (third row) and a real \ac{nisq} device (fourth row) for digit from $1$ to $9$. We also report the \acs{fid} metric values between the original full training set and a same-sized generated dataset.} 
    }
    \label{fig:digitEvolution_sim_1_9-samples}
\end{figure*}

In \cref{fig:digitEvolution_sim}, we show forward and backward processes for image generation implemented by the simulator \emph{fake\_brisbane} for the digit $0$ of the MNIST dataset. \change{To integrate the results of \cref{fig:qswEvolution,fig:digitEvolution,fig:digitEvolution_sim} to the other digits,} in \cref{fig:digitEvolution_sim_1_9-samples} we illustrate the generated images for the digits $1, 2, \dots, 9$ of MNIST obtained with a hybrid \ac{qsw} dynamics ($\omega = 0.3$), the simulator \emph{fake\_brisbane}, and the real IBM quantum hardware \emph{ibm\_brisbane}. \change{All the models used for the generation of these digits are trained on disjoint subsets of MNIST containing only the specific digit.}

\section*{Acknowledgements}
M.P.,~S.M.~and~F.C. acknowledge financial support from the PNRR MUR project PE0000023-NQSTI. S.M~and~F.C. also acknowledge financial support from Fondazione Cassa di Risparmio di Firenze.

\section*{Author Contributions}
M.P.~and S.M.~performed the implementation and experiments. 
M.P., S.M., F.A.V.~and F.C.~discussed and analyzed the results. 
M.P., S.M., F.C.~conceived the methodology, while F.C.~proposed and supervised the project.
M.P., S.M.~and F.A.V.~wrote the original Draft. 
M.P., S.M., F.A.V.~and F.C.~performed the review and Editing.

\section*{Competing Interests}
The authors declare no competing interests.

\section*{Availability of data and materials}
All data used in the manuscript are
publicly available in the referenced repositories. Code from the authors is available upon request.

\clearpage
\clearpage
\bibliography{bibliography}
\end{document}